\documentclass[conference]{IEEEtran}
\IEEEoverridecommandlockouts
\usepackage{cite}
\usepackage{amsmath,amssymb,amsfonts}
\usepackage{algorithmic}
\usepackage{graphicx}
\usepackage{textcomp}
\def\BibTeX{{\rm B\kern-.05em{\sc i\kern-.025em b}\kern-.08em
    T\kern-.1667em\lower.7ex\hbox{E}\kern-.125emX}}
\usepackage{amsmath,amssymb,amsfonts}
\usepackage{algorithmic}
\usepackage{graphicx}
\usepackage{subcaption}
\usepackage{textcomp}
\usepackage{multirow}
\usepackage[table,xcdraw]{xcolor}
\usepackage{soul,color}
\usepackage{hyperref}
\usepackage{todonotes}

\begin{document}

\title{Wait or Not to Wait: Evaluating Trade-Offs between Speed and Precision in Blockchain-based Federated Aggregation
\thanks{* These authors have equal contribution}
}

\author{\IEEEauthorblockN{Huong Nguyen*, Tri Nguyen*, Lauri Lovén, Susanna Pirttikangas}
\IEEEauthorblockA{\textit{Center for Ubiquitous Computing}, \textit{University of Oulu, Finland}\\
\{firstname.lastname\}@oulu.fi}
}

\maketitle
\begin{abstract}
This paper presents a fully coupled blockchain-assisted federated learning architecture that effectively eliminates single points of failure by decentralizing both the training and aggregation tasks across all participants. Our proposed system offers a high degree of flexibility, allowing participants to select shared models and customize the aggregation for local needs, thereby optimizing system performance, including accurate inference results. Notably, the integration of blockchain technology in our work is to promote a trustless environment, ensuring transparency and non-repudiation among participants when abnormalities are detected. To validate the effectiveness, we conducted real-world federated learning deployments on a private Ethereum platform, using two different models, ranging from simple to complex neural networks. The experimental results indicate comparable inference accuracy between centralized and decentralized federated learning settings. Furthermore, our findings indicate that asynchronous aggregation is a feasible option for simple learning models. However, complex learning models require greater training model involvement in the aggregation to achieve high model quality, instead of asynchronous aggregation. With the implementation of asynchronous aggregation and the flexibility to select models, participants anticipate decreased aggregation time in each communication round, while experiencing minimal accuracy trade-off. 
\end{abstract}


\section{Introduction}


Machine learning approaches are in high demand across various domains, driven by the data's rapid growth. However, traditional machine learning often requires data to be transferred to a centralized computation for a robust model formation, leading to communication costs, privacy leakage, and security concerns. In response, Federated Learning (FL)~\cite{mcmahan_2017} aims to reduce the dependence of machine learning approaches on centralized computation. 

In more detail, FL allows an array of participants to partake in training while preserving privacy through distributed training tasks. Essentially, instead of transferring data to a centralized computation, FL distributes the training tasks to various participants for local model training. Once trained, local models are sent to a server for aggregation into a global model, which is consequently shared with the participants to update their local models accordingly. By using FL, challenges associated with communication costs, data privacy, and security concerns can be mitigated. 
This characteristic of FL, along with its potential to maintain data privacy, significantly contributes to making it an attractive approach for modern computing paradigms.
However, the conventional FL still encounters several issues, including single-point failure, malicious participants, unfair incentives, local model verification, and authorization. 

Previous works have proposed solutions that address these issues by leveraging blockchain technology. Blockchain offers a trustless environment by means of a decentralized, immutable database among participants. As an example, to avoid single-point failure, the central aggregator can be shifted to a network of participants in a decentralized area~\cite{Shayan_2021_Biscotti,Awan_2019_Poster}. Another solution involving blockchain is to verify local model updates before aggregating them to the global model, which can be auditable via evidence in blockchain~\cite{Awan_2019_Poster,Peng_2022_VFChain}. To encourage participants in training model tasks, some researchers have proposed using incentive strategies based on blockchain technology \cite{Xu_2023_BESIFL,machuan_2022,Peng_2022_VFChain}. While these works have shown promise in addressing some FL issues, the intersection of blockchain and FL has raised other problems. These include trade-offs between system performance and inference accuracy, particularly when using asynchronous communication for aggregation in different scenarios, such as centralized and decentralized settings. Also, most of previous works focus on permissioned blockchain like Hyperledger Fabric\footnote{https://www.hyperledger.org/projects/fabric} and blockchain-based simulation where the system is closed and lacks flexibility or dynamicity. Hence, a real deployment for permissionless blockchain is an open question.

This work presents a feasible architecture and deployment for blockchain-based FL that addresses centralized FL issues. 
The offered architecture eliminates the aggregator as a central entity in traditional FL and instead allows the aggregation task to be deployed and maintained by any participants in the system. 
This architecture transforms FL from a traditional to a decentralized environment, with blockchain technology leveraged as the primary solution. 
Furthermore, by enabling the selection of models for personalized aggregation, our method effectively prevents the integration of poisoning (intended) or noisy models (unintended), which can lead to lower accuracy in future aggregations.
Therefore, this architecture facilitates asynchronous communication, allowing aggregation, called asynchronous aggregation, to proceed without waiting for all local training models to be completed. 
Compared to prior related works, which primarily consisted of surveys and limited experimentation, our research uniquely focuses on investigating the impact of model complexities on the performance of asynchronous aggregation in blockchain-based FL. 
Also, our work conducted on a private Ethereum aims to uncover potential challenges that may arise in real-world deployments, issues that simulations cannot fully capture. 
The decision to conduct a real deployment allows us to bridge the gap between theoretical simulations and practical implementation, providing valuable insights into the question: \textit{``How do different model complexities influence the performance of blockchain-based FL aggregation?''}


The experimental results demonstrate the feasibility of our proposal in simple learning models using asynchronous aggregation. However, complex learning models require a larger number of training models to be involved in aggregation to achieve high accuracy. Additionally, the experiments reveal a notable similarity in inference accuracy between centralized and decentralized FL settings.  

The remaining of the paper is structured as follows: Section~\ref{sec:related} provides an overview of the background and related work. Section~\ref{sec:proposed} introduces the proposed scheme, while Section~\ref{sec:experiment} presents experimental details and results. Finally, Section~\ref{sec:conclusion} summarizes this work.
\section{Background and Related Work} \label{sec:related}

\subsection{Backgrounds}
\subsubsection{System models of blockchain-leveraged Federated Learning}
One of the interesting considerations is the harmonization of architecture between FL and blockchain architecture. In detail, initial FL is introduced with a ``master-slave'' architecture, while decentralization is the foundation of blockchain technology. Regarding survey works~\cite{wang_2021,Zhu_2023_survey}, FL leverages the blockchain in three directions from the architecture perspective. 
The first view of blockchain-based FL is about \textbf{fully coupled}, which points to the FL clients handling blockchain and aggregator tasks. 
\textbf{Semi-coupled} blockchain-based FL is about the deployment of FL aggregation to blockchain participants, and FL clients train and submit local models to blockchain participants for aggregation. 
\textbf{Loosely coupled} between blockchain and FL is a concept in which the blockchain provides services, for example, access management or reputation for enhancing FL features, especially from a security viewpoint.

\subsubsection{Blockchain performance} \label{sec:related-performance}

Given the variety in blockchain platforms, an initial comparative analysis by~\cite{Pongnumkul_2017_blockchainPlatform} highlights that Fabric exhibits lower latency and higher throughput than Ethereum. Furthering this research, \cite{Dinh_2018_blockchainSystem} developed BLOCKBENCH, a tool for evaluating different blockchain platforms. Results from BLOCKBENCH demonstrate that Fabric achieves superior throughput performance when compared to both Proof-of-Work (PoW) and Proof-of-Authority variants of Ethereum. Additionally, in terms of computational and memory demands, Hyperledger Fabric consumes the fewest resources. Nonetheless, the study does not provide detailed information on block and transaction capacities for each blockchain setup, particularly noting the absence of difficulty configuration details in PoW-based Ethereum. Also, the limited interactions among smart contracts pose another area of concern. Therefore, previous studies~\cite{Nguyen_2022_car,NGUYEN2023606} detailed the inverse effect of block size and throughput. Particularly, those works evaluate the system's performance in various scenarios increasing the block size, which findings are about the more number of network participants and the high capacity of messages influences the system performance, especially in throughput. Regarding the influence of system performance via the number of participants in a blockchain-based FL, Peng et al.~\cite{Peng_2022_VFChain} find a half reduction in throughput and an increase in execution time from the doubled number of participants. To boost the blockchain performance, Sethi et al.~\cite{Sethi_2024} point the usage of reinforcement learning to predict the PoW's difficulty in consensus rounds to enhance blockchain performance, especially in the usage of blockchain-based FL where the number of participants is flexible.

Due to the previous comprehensive works in blockchain performance in different scenarios, those findings related to network measurement are accepted in this work, which means the asynchronous aggregation with less communication gains achievements in comparison with the synchronous federated aggregation waiting and aggregating the entire local models.  

\subsection{Related work}

One of the initial proposals for integrating blockchain to enhance security in FL is presented in~\cite{Awan_2019_Poster}. This study introduces a semi-coupled blockchain-based FL framework that leverages blockchain technology to prevent privacy leaks. Specifically, it utilizes blockchain to manage data flows and implement a verification mechanism that supports a fully decentralized environment. However, the experimental setup described is relatively simple, involving a single blockchain node, and it does not address the impact of complex models on asynchronous aggregation.

Flchain~\cite{Majeed_2019} also explores a semi-coupled blockchain-based FL system, designed to bolster the security of FL through the creation of channels for learning multiple global models. In this system, local devices first train their models independently before exchanging them via blockchain to form a cohesive global model. This approach allows devices the flexibility to select channels or global models that best suit their training needs. 
Despite its innovative approach, this early work on blockchain integration in FL lacks detailed experimental validation and comprehensive settings description.

Lu et al.~\cite{Lu_2020_BCFL} discuss the application of blockchain combined with FL to ensure privacy-preserving data sharing within the industrial IoT. A permissioned blockchain is employed to manage the computation and storage of FL models, facilitating the auditing of data usage permissions. In other words, \cite{Lu_2020_BCFL} is a semi-coupled blockchain-based FL for a marketplace which deals with the privacy-preserving aspect in sharing data from data holders. 
While this study builds upon a promising concept and conducts experiments with two datasets, it falls short in extensively exploring the effects of varying model complexities or the practical deployment challenges encountered in blockchain-based FL systems.

Kang et al.~\cite{kang_2020} proposed a blockchain-empowered federated edge learning. The study employs a dual blockchain structure: a main public blockchain for global models and consortium sub-chains for local models from edge servers. This setup supports decentralization while catering to real-time applications. They introduced a ``Proof-of-Verifying'' consensus to filter malicious updates, focusing on gradient compression to improve communication efficiency. This consensus evaluates local updates against a testing dataset using a threshold, ensuring only valid updates contribute to the global model.

From FL's aggregation aspect, Ramanan et al.~\cite{ramanan_2020}'s BAFFLE introduces a novel approach to FL by employing blockchain technology to facilitate aggregator-free operations. It leverages smart contracts for the aggregation of the global model, eliminating the need for a central aggregator. 
Despite its innovative approach to decentralizing model aggregation and enhancing security, BAFFLE faces significant challenges with increased computation and communication costs. The system incorporates smart contracts for interactions and updates in the blockchain, ensuring a seamless blend of local training model updates and global model aggregation.
Similar to BAFFLE, PIRATE~\cite{Zhou_2020} utilizes blockchain technology for decentralized learning, storing each community's global model on separate blockchains. This facilitates inter-community model updates. PIRATE addresses Byzantine attacks, which can occur during model parameter updates and gradient aggregations, by enhancing the system's resilience against such threats in distributed learning environments.

Li et al.~\cite{Li_2021} introduce a blockchain-based FL framework named BFLC, which utilizes a committee consensus mechanism to enhance both efficiency and security. This framework employs blockchain technology for securely storing global models and exchanging local model updates. BFLC improves performance through a committee selection consensus mechanism and uses smart contracts for autonomously aggregating global models, ensuring that only authorized devices can contribute updates. This approach significantly enhances the security and efficiency of FL systems.

Shayan et al.~\cite{Shayan_2021_Biscotti} leverage blockchain and cryptographic techniques for privacy-preserving machine learning process through a semi-coupled blockchain-based FL. Their system, Biscotti, is noted for its resilience to fault-tolerant and malicious adversaries. In Biscotti, each blockchain's block represents a training iteration, in which a selected committee aggregates and verifies local updates before issuing rewards to trainers. These rewards, such as reputed scores, help in the selection of committee members for aggregating models in FL and gaining consensus among blockchain participants. 

Peng et al.~\cite{Peng_2022_VFChain} present VFChain, a semi-coupled blockchain-based FL for auditability and verifiability. A selected committee aggregates models alongside verifiable proofs recorded on blockchain for auditability. The committee, selected from a reputation system, receives local gradients from participants for aggregation.
The experiments are based on a centralized Fabric platform and a convolution neural network, and include mechanisms for verifying global models through voting procedures; however, it does not support personalized aggregation in the aggregation process, limiting its flexibility.
Despite enabling decentralized features, those previous works face other issues regarding performance and decentralized issues. 
BLADE-FL~\cite{machuan_2022} builds on the integration of blockchain to enhance FL, aiming to overcome efficiency limitations seen in prior works. It emphasizes privacy, resource allocation, and addressing free-rider issues, distinguishing itself by requiring specific contributions for maintaining the blockchain and local model training. BLADE-FL introduces a mechanism for assigning training tasks and rewards, addresses participant anonymity by adding random noise to data (though it affects learning performance), and proposes a method using pseudo-noise sequences to identify and penalize non-contributing participants.

Regarding the communication aspect, Wilhelmi et al.~\cite{Wilhelmi_2022} explore blockchain's role in decentralizing FL to form semi-coupled blockchain-based FL focusing on the freshness of model updates via the ``age-of-block'' metric to measure time differences between local model updates. This concept helps analyze how blockchain's real-world application affects learning accuracy, addressing potential inconsistencies and network delays, especially blockchain fork issues. Their findings highlight the FL framework's resilience in maintaining accuracy across various environmental configurations. 

Xu et al.~\cite{Xu_2023_BESIFL} introduce BESIFL, a loosely coupled blockchain-based FL, to prevent malicious nodes through an incentive mechanism. BESIFL aims to enhance inference accuracy by selecting credible nodes using a reputation-based incentive system. The BESIFL employs a credit-based approach for node selection, distinguishing training and verification participants through two specific scores derived from a contribution-based incentive mechanism. Via token-based reward and punishment, smart contract-based model demander playing a role as the FL server for aggregation offers reward or punishment training or malicious participants. This incentive strategy enhances the accuracy compared to previous models.
The experiments are deployed in Hyperledger Fabric with a centralized server with a Kafka setting, and IPFS is utilized for decentralized storage, with hash data and reputation scores stored on the blockchain for verification.  
Despite these advancements, the study does not explore how varying complexities of models might affect inference accuracy, which remains an open area.

Finally, while prior works have shown promise in addressing certain FL challenges, the majority of the existing literature focuses on permissioned blockchains such as Hyperledger Fabric, or blockchain-based simulations with closed systems lacking flexibility and dynamism. 
A prevailing weakness among prior studies is also the inadequate evaluation of their systems' compatibility across a spectrum of model complexities.
This leaves a significant gap in exploring real deployments for permissionless blockchains, which is exactly where our work delves into the intricate details of this intersection under multiple settings, particularly the accuracy comparison between centralized and decentralized settings along with the variety of model complexities.

\section{fully coupled blockchain-based Federated Learning} \label{sec:proposed}

    \subsection{Decentralization via Blockchain-based FL}
        \begin{figure*}[!]
            \centering
                \includegraphics[width=\textwidth]{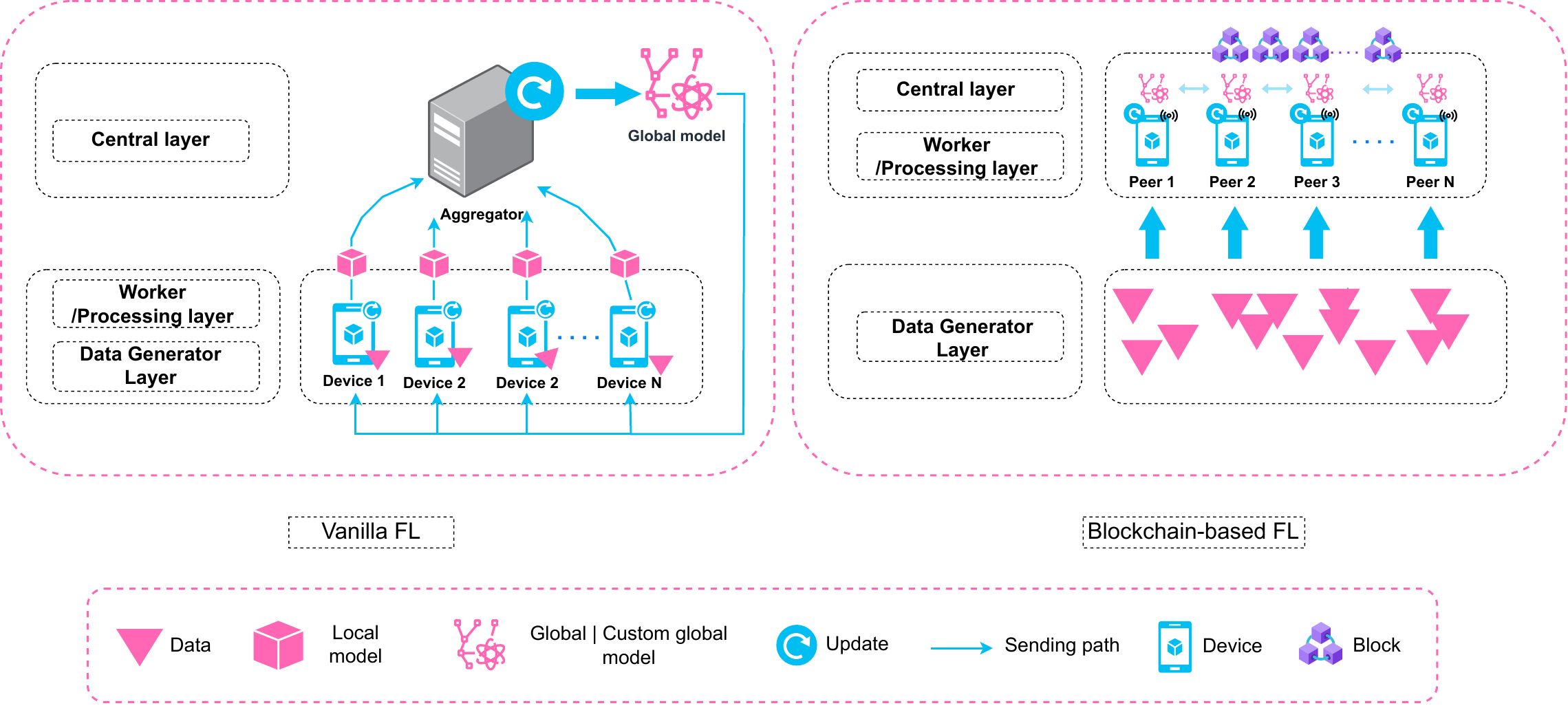}
            \caption{Vanilla vs proposed FL architecture illustrations}
            \label{fig:architecture}
        \end{figure*}
       Due to single-point failure from traditional FL, leveraging blockchain technology regarding fully coupled integration is a promising solution for shifting to a decentralized setting. 
       Following traditional FL, the aggregation duty as centralization is to form a global model based on an array of local models; however, from the vulnerable perspective, single-point failure is a downside, and the transformation to decentralization is a promising solution.
       In particular, Figure~\ref{fig:architecture} illustrates the comparison between Vanilla and decentralized learning. Based on the prior works, we found that the traditional FL systems work with two layers: the client layer, which handles data generation, self-processing, and making local models, and the aggregator layer forming the global models. 
       With the decentralized aspect, this research attempts to rearrange the architecture and roles in the system, where the data generator is separated/organized into one layer meanwhile the worker node, as well as the aggregator, are merged into one layer. 
       Consequently, we will have multiple aggregators working as peers in an asynchronous blockchain-assisted FL system, supporting in detecting nodes with abnormal models.
       With fully coupled blockchain interference, the proposed system gains a lot of perks, which we consider three prominent cases:
        \begin{itemize}
            \item Case 1 - flexibility in choosing shared models and customizing an aggregation for the local usage
            \item Case 2 - flexibility in aggregation time or asynchronous aggregation
            \item Case 3 - ensuring non-repudiation of the participant about their models
        \end{itemize}

        Leveraging the fact that peers will automatically share their new data with others, therefore, when we put the blockchain into an FL system, peers - those working as local workers and aggregators here will broadcast their local model (weights) to the rest of the participants in the network. As a consequence, each peer will have all the shared models from others and is expected to have the ability to choose ones to join its customized aggregation process freely or asynchronous aggregation. 
        To do this, a test dataset is prepared to evaluate the fitness of the shared model. If the evaluation is over a pre-set threshold,  the worker will then include that model in their aggregation process; otherwise, it will be ignored.  This will help when the updated model is too general and contains weights from the outliers due to the heterogeneous data from other regions or scopes, leading to an accuracy decrease.

        \subsection{Asynchronous Aggregation in Decentralized Learning}
        Due to the large communication from the bottleneck, the proposed learning system will be an asynchronous system, where there is no need to wait for all participants in the network to complete their local training to start the aggregation progress. 
        Working in the conventional method, where each peer awaits all other peers' updates for aggregation will create a heavier computational load for workers. This becomes more observable when we have millions of participants in the network and each has to evaluate the relevance and suitability of numerous shared models before starting their customized aggregation. From that, bottlenecks can be formed at peer-level and the performance is then reduced. 
        Therefore, each aggregator can desire how many local updates she/he would use to aggregate instead of waiting for total update models. However, the issue from this consideration leads to local generalization and reduces the knowledge from the global. 
        Based on this analysis, we come up with a contemplation that varying complexities in models may impact the inference accuracy in the asynchronous aggregation processes of blockchain-based federated learning systems.


        Regarding the flexible configuration protocol and dynamic participants, the system can operate in either decentralized learning with various customized models or FL regarding the identical global model. In the first case, the system can go as decentralized learning which points to the freedom of customizing the current peer's model with an arbitrary number of local models to satisfy the peer's specific test dataset. Meanwhile, the second method supports gaining a global viewpoint from the entire network's peers to vote and support a bunk of local models regarding a consensus like a global model; however, instead of a fixed single aggregator, this mechanism allows any peer to become the aggregator to avoid the single-point failure.

        \subsection{Hierarchy of blockchain-based FL}
        For a sophisticated blockchain-based FL design, we deliberate the adoption of a middle layer by which the data is transferred to the closest local server instead of traditional FL, which requires training in data generation before sending it to a centralized aggregation server. 
        Regarding the lack of data in training, the local server gathers data nearby before training a local model.
        In detail, local servers communicate and exchange their local models after training. At this stage, the peers will have two options for aggregation: (1) customizing an arbitrary number of local models or (2) agreeing on a common block of local updates as representing the potential global update for the next state. 
        Besides, the usage of this layer enables the intelligence at the edge that has been discussed as the most interesting topic from various communities~\cite{Peltonen_2022}, regarding the cloud-edge continuum, due to its availability and performance.

\section{Evaluation} \label{sec:experiment}

    \subsection{Setup and Environment}

    \subsubsection{Blockchain Configuration}
\begin{figure}
        \centering
        \includegraphics[width=0.47\textwidth]{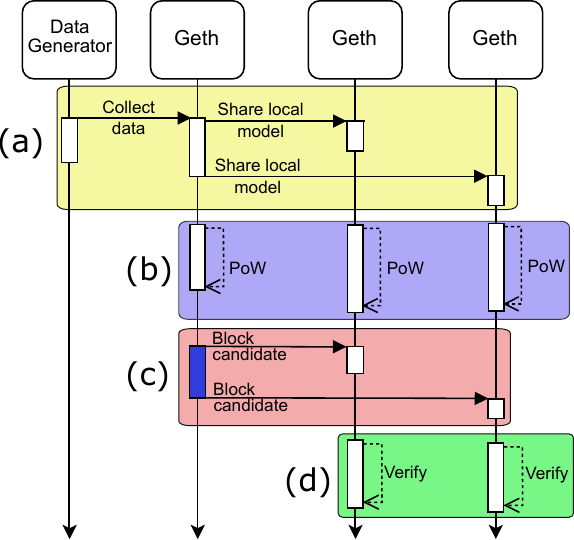}
        \caption{Ethereum  setting for the experiment via three peers deploying Ethereum application denoted Geth with different tasks: (a) Interaction between data generator and blockchain participants, (b) PoW consensus to select a leader, (c) Block formation, (d) Block validation}
        \label{fig:ethereum}
    \end{figure}

    We decided to utilize the Ethereum platform\footnote{https://ethereum.org/en/} to deploy our proposal. The main idea with Ethereum is a well-known permissionless blockchain, which should fit our proposal as aforementioned. Due to the permissionless blockchain platform, the computation cost from PoW consensus cannot be avoided; however, Ethereum enables openness in collaboration learning. The workflow of Ethereum can be viewed in Figure~\ref{fig:ethereum}. In the beginning, from the yellow part of the figure, the data is collected from the data generator for training before sharing it with other participants in the blockchain network. Parallelly, blockchain participants execute a PoW procedure to determine the leader of the current consensus round in the blue region of Figure~\ref{fig:ethereum}. Once the leader is determined, a potential block is broadcast from the leader for verification from other participants in regions red and green, respectively.

    The blockchain experiments are based on the Ethereum framework which setup encompassed three virtual entities, called Geth:1.10.12 along with Truffle:5.4.28\footnote{https://archive.trufflesuite.com/}, on a physical machine: an Intel(R) Core(TM) i7-8700 CPU @ 3.20 GHz, 3.19 GHz, 32 GB RAM, and 64-bit OS x64-based processor. Concretely, based on the VirtualBox tool, each virtual machine resource has 6 GB RAM and 4 processors. The asynchronous aggregation is based on a solidity-language smart contract. 
    Regarding the interaction between FL local model training and blockchain-based smart contract, we utilize web3\footnote{https://web3js.readthedocs.io/en/v1.2.1/getting-started.html} library and asynchronous communication via NodeJS for propagating local models and collecting aggregated one for the next iterator. 
    Furthermore, building upon the system performance findings from previous works discussed in Section~\ref{sec:related-performance}, we have configured Ethereum without block size and transaction size constraints. Specifically, we ensure that the transaction size exceeds the model's size, as detailed in the gas conversion aspect outlined in~\cite{NGUYEN2023606}. This adjustment accounts for Ethereum's complexity, which is measured in units of gas.

    

    
    \subsubsection{Federated Learning Configuration}
    Regarding the training, we come up with three considerations that need to be decided, including dataset, models, and training. 
    
    \textbf{Dataset:} In determining the selection criteria for these datasets, we conducted a thorough review of prior works pertaining to FL and found that MNIST, CIFAR-10, and CIFAR-100 are commonly employed as benchmark datasets. We then opted to use CIFAR-10 due to its colored images and the limited number of classes (ten). This decision was made in contrast to MNIST, which contains only binary images, and CIFAR-100, with a significantly larger number of labels (one hundred) while maintaining the same dataset size. The CIFAR-10 dataset consists of 60000 images in the size of 32x32, number of classes in total is 10 with a variety of labels on animals (bird, cat, deer, dog, frog, horse) and vehicles (airplane, automobile, ship, truck).

    \textbf{Model:} In the model selection phase, we considered running our experiment on multiple models, ranging from simple Neural Network (Simple NN) architectures to more complex ones, such as EfficientNet, developed by Google. Specifically, we have opted to work with both Simple NN model and the most resource-efficient variant of EfficientNet, denoted as EfficientNet-B0.
    
    

    \textbf{Training:} We use PyTorch to implement the training with chosen models: Simple NN and EfficientNet-B0. The Simple NN model is constructed from scratch with only 62K parameters and approximately 248KB in size, whereas, for EfficientNet-B0 (parameters count 5.3M, size 21.2MB), we employ transfer learning by modifying its final layer. FedAvg\cite{mcmahan_2017} algorithm is utilized for model aggregation during our FL training process.


    \subsection{Experiments and Results}
    
    \subsubsection{Experiments} 

    
    Regarding the experiments, we illustrate two scenarios of deploying FL in centralized and decentralized environments. In both cases, we consider to assess the fitness of the updated model. Specifically, Section~\ref{sec:proposed} points out the usage of test data from each participant to evaluate the receiving model updates before deciding to aggregate at each communication round.
    
    \textbf{Centralized setting:}
    In the centralized scenario, we deploy a Vanilla FL with ten communication rounds in which each training device trains five epochs for local updates before sending for aggregation. In detail, three virtual machines representing three local devices A, B, and C, are used for training local models before sharing them with the aggregator to aggregate the global model. This global is then shared back to local devices for the next round.
    Besides, we present two types of aggregation from the aggregator, called ``consider'' and ``not consider''. ``not consider'' is Vanilla's traditional approach, where the aggregator collects and aggregates all three local updates. Meanwhile, in the ``consider'' case, the aggregator chooses the best combination of local models via a default test set.

    \textbf{Decentralized setting:}
    In the decentralized scenario, we deploy a blockchain-based FL with Ethereum. In particular, three virtual machines implement both local training and aggregation tasks, instead of using a specific aggregator for global aggregation. Therefore, each device's results vary based on the diversity of aggregation. For the next training round, the combined model with the highest accuracy from the last round is utilized. As a result, the chosen model combination may vary between rounds. In cases where multiple combinations achieve the same performance or accuracy, the device selects one of them randomly. We also keep the same FL training as Vanilla's setting, which is ten communication rounds and five epochs for training at local devices.

    \subsubsection{Results} 
    Regarding the experimental results, we investigate the similarity as well as the main difference findings in Vanilla and blockchain-based FL settings. 
    
\begin{figure*}[h!]
       \begin{subfigure}{\textwidth}
       \centering
           \includegraphics[scale=.3]{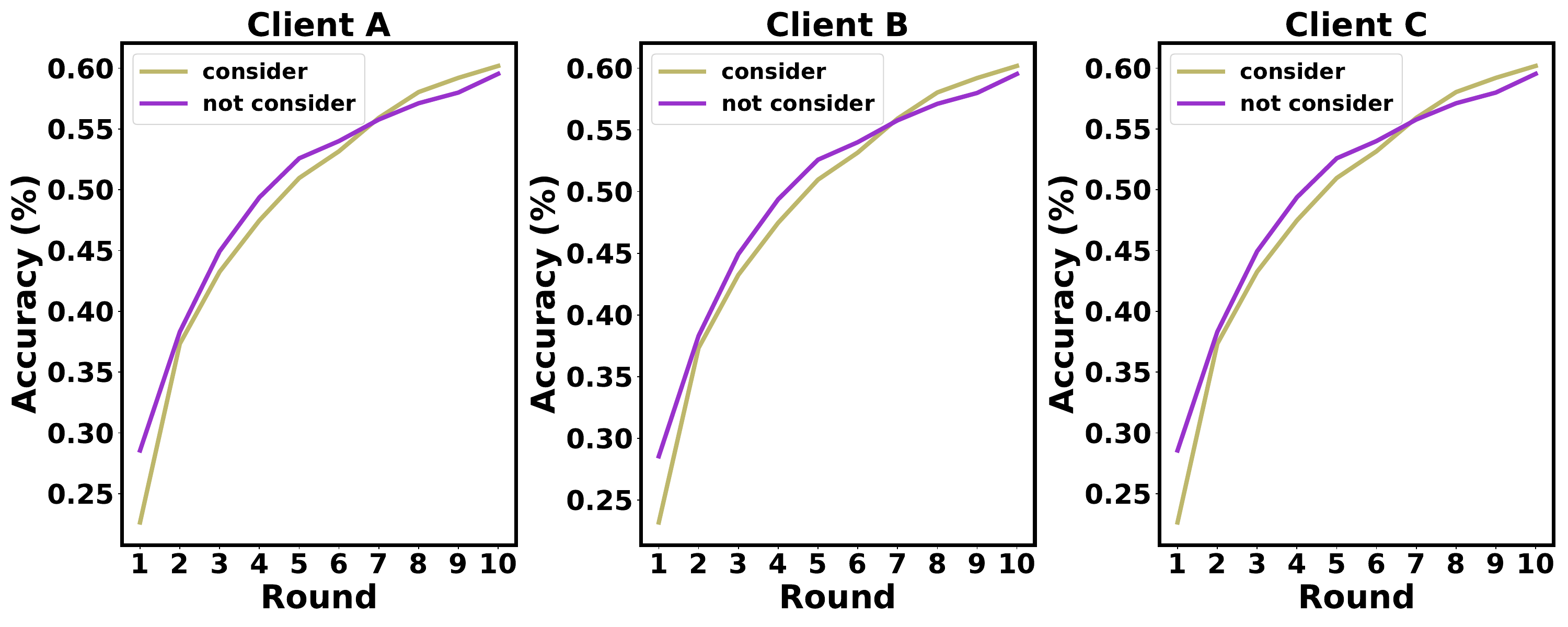} 
           \caption{SimpleNN}\label{fig:consider-non-simplenn}
       \end{subfigure}\hfill
       \begin{subfigure}{\textwidth}
       \centering
           \includegraphics[scale=.3]{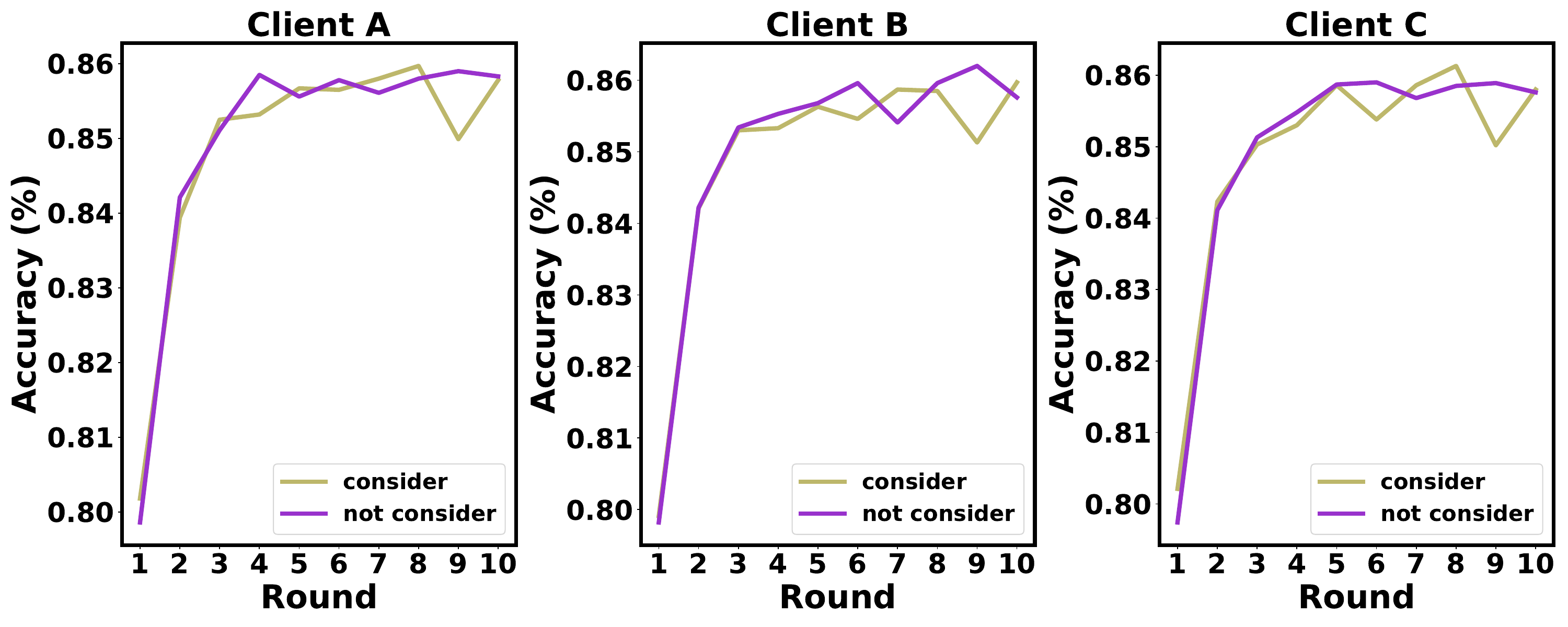}
           \caption{Efficient B0}\label{fig:consider-non-b0}
       \end{subfigure}\hfill
       \caption{Vanilla FL: Test accuracy (\%) on two aggregation types}
       \label{fig:consider-vs-non}
    \end{figure*}

    \textbf{Vanilla setting:}
     In Table~\ref{tab:vani-acc}, we report the accuracy values of participants in each communication round, considering two aggregation types at the aggregator: ``consider'' and ``not consider''. These values are measured by evaluating the aggregated model's performance on the test data from individual local clients. We also include Figure~\ref{fig:consider-vs-non} to visualize the reported values. 
     Regarding comparative results, we observe that having the option to select models at the aggregator (denoted as ``consider'') generally results in better accuracy for participating clients, especially when training simple models. The difference here is minor, at only 0.65\%, between 59.53\% (``not consider'') and 60.18\% (``consider''). On the other hand, when dealing with more complex models, we achieve higher performance (around 85-86\% across clients), accompanied by some fluctuations between these two aggregation types, as depicted in Figure~\ref{fig:consider-non-b0}.
     Those results can point out the similarity in the accuracy despite the keeping or discarding of imperfect models in the two aggregation approaches. Due to this finding, the blockchain-based FL is expected to be a potential solution for unfit or abnormal model detection in the aggregation process.
    
    
    \begin{table*}[]
    \caption{Vanilla FL: Clients' test accuracy (\%) on two aggregation types}
    \label{tab:vani-acc}
    \centering
    \def\arraystretch{1.1}
    \begin{tabular}{|c|c|c|cccccccccc|}
    \hline
    \textbf{Model} &
      \textbf{Client} &
      \textbf{Params} &
      \multicolumn{10}{c|}{\textbf{round}} \\ \hline
     &
       &
       &
      \multicolumn{1}{c|}{\textbf{1}} &
      \multicolumn{1}{c|}{\textbf{2}} &
      \multicolumn{1}{c|}{\textbf{3}} &
      \multicolumn{1}{c|}{\textbf{4}} &
      \multicolumn{1}{c|}{\textbf{5}} &
      \multicolumn{1}{c|}{\textbf{6}} &
      \multicolumn{1}{c|}{\textbf{7}} &
      \multicolumn{1}{c|}{\textbf{8}} &
      \multicolumn{1}{c|}{\textbf{9}} &
      \textbf{10} \\ \hline
    \multirow{6}{*}{Simple NN} &
      \multirow{2}{*}{A} &
      Consider &
      \multicolumn{1}{c|}{0.2263} &
      \multicolumn{1}{c|}{0.3733} &
      \multicolumn{1}{c|}{0.4325} &
      \multicolumn{1}{c|}{0.4747} &
      \multicolumn{1}{c|}{0.5096} &
      \multicolumn{1}{c|}{0.5316} &
      \multicolumn{1}{c|}{0.5591} &
      \multicolumn{1}{c|}{0.5803} &
      \multicolumn{1}{c|}{0.592} &
      \textbf{0.6018} \\ \cline{3-13} 
     &
       &
      Not consider &
      \multicolumn{1}{c|}{0.2855} &
      \multicolumn{1}{c|}{0.3831} &
      \multicolumn{1}{c|}{0.4493} &
      \multicolumn{1}{c|}{0.4937} &
      \multicolumn{1}{c|}{0.5258} &
      \multicolumn{1}{c|}{0.54} &
      \multicolumn{1}{c|}{0.5577} &
      \multicolumn{1}{c|}{0.5711} &
      \multicolumn{1}{c|}{0.5799} &
      0.5953 \\ \cline{2-13} 
     &
      \multirow{2}{*}{B} &
      Consider &
      \multicolumn{1}{c|}{0.2319} &
      \multicolumn{1}{c|}{0.3733} &
      \multicolumn{1}{c|}{0.4325} &
      \multicolumn{1}{c|}{0.4747} &
      \multicolumn{1}{c|}{0.5096} &
      \multicolumn{1}{c|}{0.5316} &
      \multicolumn{1}{c|}{0.5591} &
      \multicolumn{1}{c|}{0.5803} &
      \multicolumn{1}{c|}{0.592} &
      \textbf{0.6018} \\ \cline{3-13} 
     &
       &
      Not consider &
      \multicolumn{1}{c|}{0.2855} &
      \multicolumn{1}{c|}{0.3831} &
      \multicolumn{1}{c|}{0.4493} &
      \multicolumn{1}{c|}{0.4937} &
      \multicolumn{1}{c|}{0.5258} &
      \multicolumn{1}{c|}{0.54} &
      \multicolumn{1}{c|}{0.5577} &
      \multicolumn{1}{c|}{0.5711} &
      \multicolumn{1}{c|}{0.5799} &
      0.5953 \\ \cline{2-13} 
     &
      \multirow{2}{*}{C} &
      Consider &
      \multicolumn{1}{c|}{0.2263} &
      \multicolumn{1}{c|}{0.3733} &
      \multicolumn{1}{c|}{0.4325} &
      \multicolumn{1}{c|}{0.4747} &
      \multicolumn{1}{c|}{0.5096} &
      \multicolumn{1}{c|}{0.5316} &
      \multicolumn{1}{c|}{0.5591} &
      \multicolumn{1}{c|}{0.5803} &
      \multicolumn{1}{c|}{0.592} &
      \textbf{0.6018} \\ \cline{3-13} 
     &
       &
      Not consider &
      \multicolumn{1}{c|}{0.2855} &
      \multicolumn{1}{c|}{0.3831} &
      \multicolumn{1}{c|}{0.4493} &
      \multicolumn{1}{c|}{0.4937} &
      \multicolumn{1}{c|}{0.5258} &
      \multicolumn{1}{c|}{0.54} &
      \multicolumn{1}{c|}{0.5577} &
      \multicolumn{1}{c|}{0.5711} &
      \multicolumn{1}{c|}{0.5799} &
      0.5953 \\ \hline \hline
    \multirow{6}{*}{Efficient-B0} &
      \multirow{2}{*}{A} &
      Consider &
      \multicolumn{1}{c|}{0.8018} &
      \multicolumn{1}{c|}{0.8394} &
      \multicolumn{1}{c|}{0.8525} &
      \multicolumn{1}{c|}{0.8532} &
      \multicolumn{1}{c|}{0.8567} &
      \multicolumn{1}{c|}{0.8565} &
      \multicolumn{1}{c|}{0.858} &
      \multicolumn{1}{c|}{0.8597} &
      \multicolumn{1}{c|}{0.8499} &
      0.8578 \\ \cline{3-13} 
     &
       &
      Not consider &
      \multicolumn{1}{c|}{0.7986} &
      \multicolumn{1}{c|}{0.8421} &
      \multicolumn{1}{c|}{0.8511} &
      \multicolumn{1}{c|}{0.8585} &
      \multicolumn{1}{c|}{0.8556} &
      \multicolumn{1}{c|}{0.8578} &
      \multicolumn{1}{c|}{0.8561} &
      \multicolumn{1}{c|}{0.858} &
      \multicolumn{1}{c|}{0.859} &
      \textbf{0.8583} \\ \cline{2-13} 
     &
      \multirow{2}{*}{B} &
      Consider &
      \multicolumn{1}{c|}{0.7991} &
      \multicolumn{1}{c|}{0.8421} &
      \multicolumn{1}{c|}{0.853} &
      \multicolumn{1}{c|}{0.8533} &
      \multicolumn{1}{c|}{0.8563} &
      \multicolumn{1}{c|}{0.8546} &
      \multicolumn{1}{c|}{0.8587} &
      \multicolumn{1}{c|}{0.8585} &
      \multicolumn{1}{c|}{0.8513} &
      \textbf{0.8597} \\ \cline{3-13} 
     &
       &
      Not consider &
      \multicolumn{1}{c|}{0.7982} &
      \multicolumn{1}{c|}{0.8422} &
      \multicolumn{1}{c|}{0.8534} &
      \multicolumn{1}{c|}{0.8553} &
      \multicolumn{1}{c|}{0.8568} &
      \multicolumn{1}{c|}{0.8596} &
      \multicolumn{1}{c|}{0.8541} &
      \multicolumn{1}{c|}{0.8596} &
      \multicolumn{1}{c|}{0.862} &
      0.8576 \\ \cline{2-13} 
     &
      \multirow{2}{*}{C} &
      Consider &
      \multicolumn{1}{c|}{0.8021} &
      \multicolumn{1}{c|}{0.8423} &
      \multicolumn{1}{c|}{0.8503} &
      \multicolumn{1}{c|}{0.853} &
      \multicolumn{1}{c|}{0.8586} &
      \multicolumn{1}{c|}{0.8538} &
      \multicolumn{1}{c|}{0.8586} &
      \multicolumn{1}{c|}{0.8613} &
      \multicolumn{1}{c|}{0.8502} &
      \textbf{0.858} \\ \cline{3-13} 
     &
       &
      Not consider &
      \multicolumn{1}{c|}{0.7974} &
      \multicolumn{1}{c|}{0.8411} &
      \multicolumn{1}{c|}{0.8513} &
      \multicolumn{1}{c|}{0.8548} &
      \multicolumn{1}{c|}{0.8587} &
      \multicolumn{1}{c|}{0.859} &
      \multicolumn{1}{c|}{0.8568} &
      \multicolumn{1}{c|}{0.8585} &
      \multicolumn{1}{c|}{0.8589} &
      0.8576 \\ \hline 
    \end{tabular}
    \end{table*}
    
\begin{figure*}[h!]
        \begin{subfigure}{\textwidth}
        \centering
           \includegraphics[scale=.22]{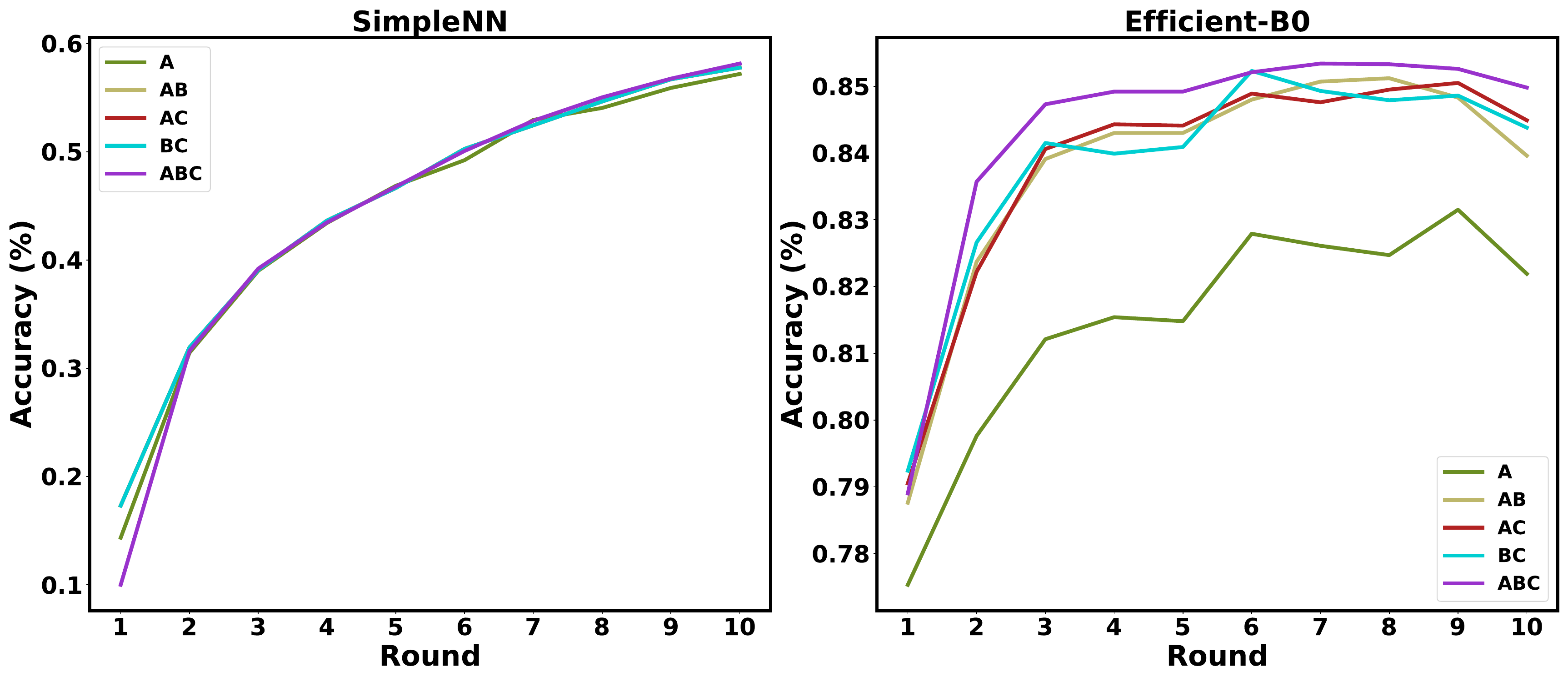} 
           \caption{Client A}
        \end{subfigure}\hfill
        \begin{subfigure}{\textwidth}
        \centering
           \includegraphics[scale=.22]{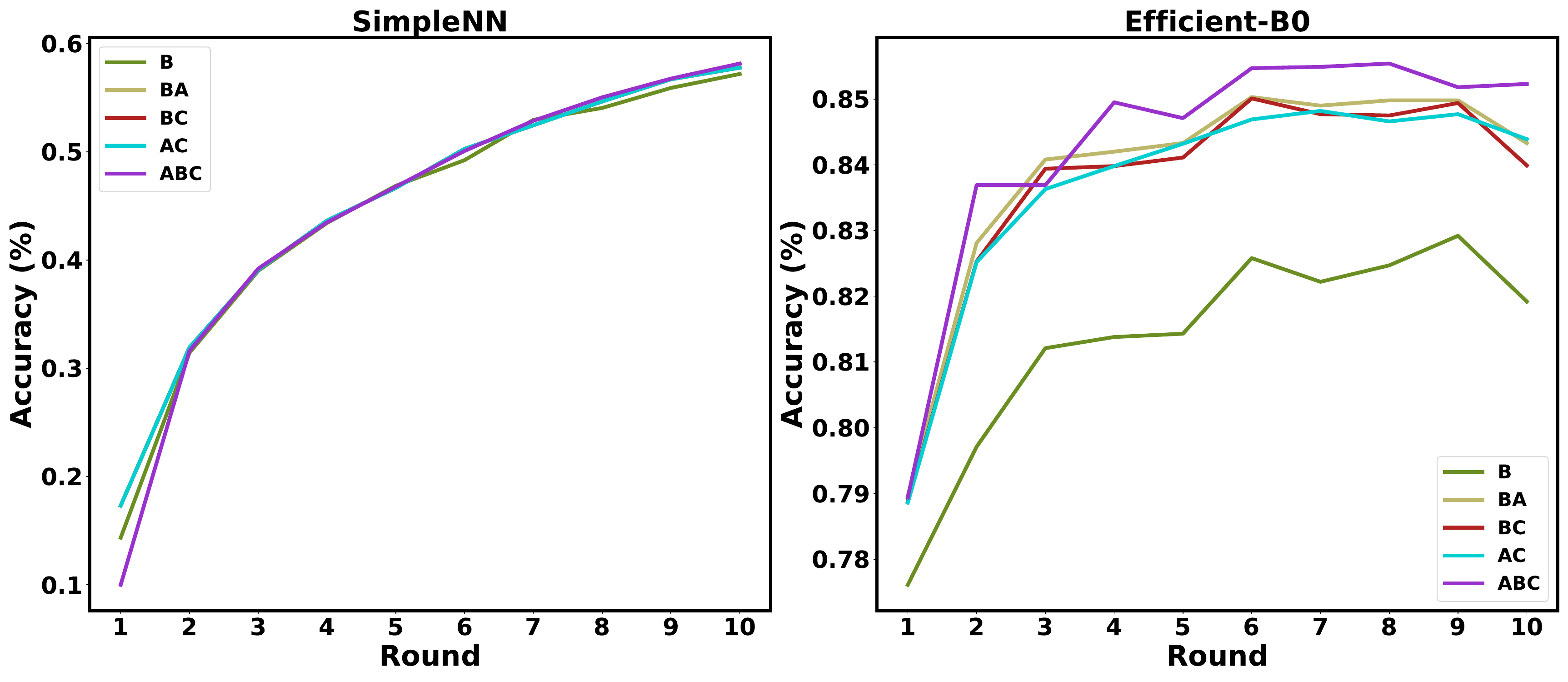}
           \caption{Client B}
        \end{subfigure}\hfill
        \begin{subfigure}{\textwidth}
        \centering
           \includegraphics[scale=.22]{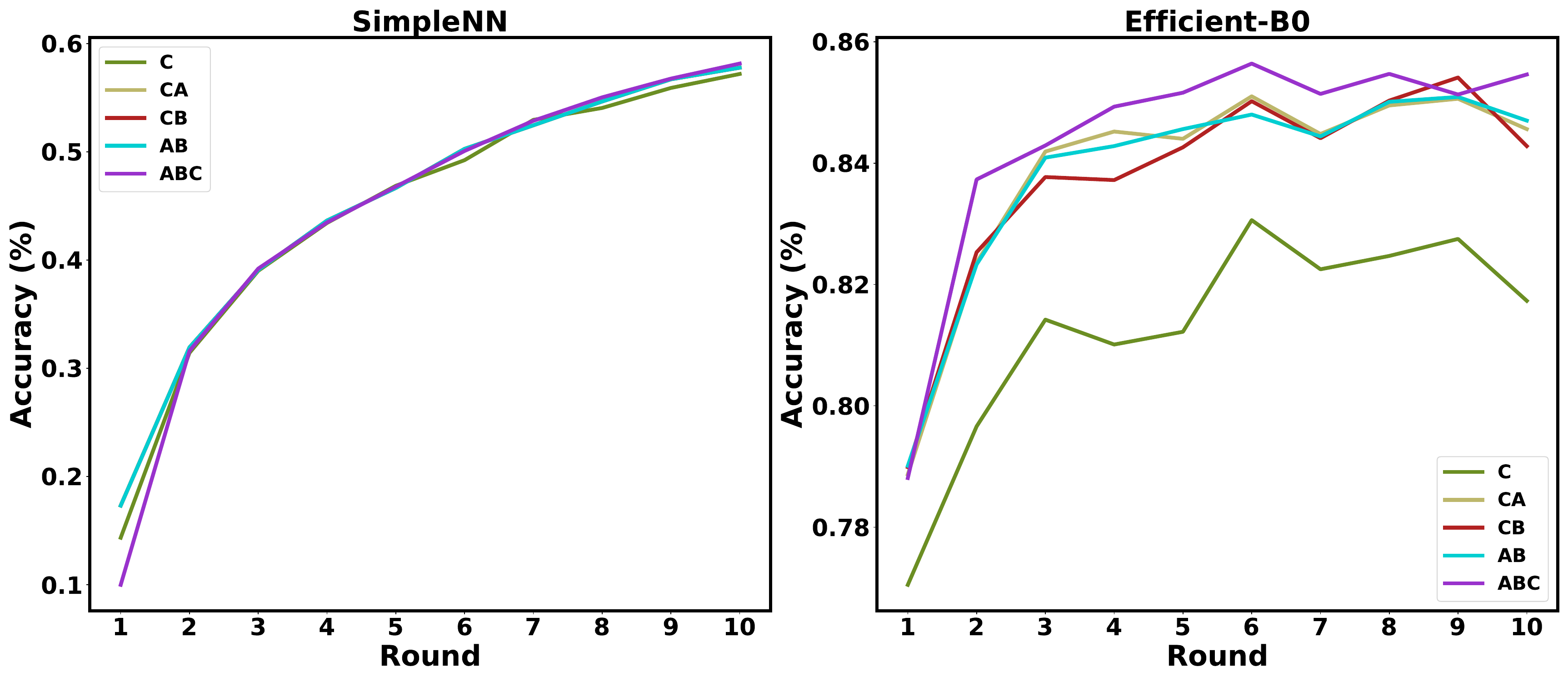}
           \caption{Client C}
        \end{subfigure}
        \caption{Blockchain-based FL: Test accuracy (\%) on different model combinations}
        \label{fig:plot-cliabc}
    \end{figure*}

    \textbf{Decentralized setting:}
     Concerning experiments of blockchain-based FL, the accuracy for each round is reported in Table \ref{tab:proposed-acc-A}, \ref{tab:proposed-acc-B}, and \ref{tab:proposed-acc-C}, corresponding to the respective device. Overall, the experiments with Simple NN models demonstrate consistent accuracy values across participants, while variations in accuracy values are observed among different devices in the case of Efficient-B0 models. In terms of model combination,  the one that has the highest aggregation accuracy varies across communication rounds, particularly for simple models. However, for a more complex model like Efficient-B0, aggregating all models consistently yields the highest accuracy in most rounds, except for specific instances: round 1 at client A with model combination BC (78.97\% vs. 78.9\% for the ABC model), round 3 at client A with combination BC (85.23\% vs. 85.21\% for ABC), round 1 at client B with model combination BA (78.97\% vs. 78.94\% for ABC), round 3 at client B with model BA (84.08\% vs. 83.69\% for ABC), round 1 at client C with model combination AB (79.01\% vs. 78.81\% for ABC), and round 9 at client C with model CB (85.41\% vs. 85.13\% for ABC). 
     As shown in Figure~\ref{fig:plot-cliabc}, the similarity of various aggregations is evident among devices in the case of Simple NN, but this does not hold for Efficient-B0. However, it is worth noting that when training complex models like Efficient-B0, it is more beneficial for participating clients to collaborate by combining their local models with others, as using solely their local models consistently results in lower or sub-optimal performance. The blockchain-based FL experiment highlights that aggregating the entire set of models in complex models yields superior results, whereas, in the case of simple models, accuracy remains relatively consistent.
     
     Consequently, a trade-off question, regarding the \textbf{aggregation time} and \textbf{model quality}, is raised: \textit{``Should we prioritize waiting for all models for aggregation, or accept a slight reduction in accuracy to expedite the process asynchronously?''}
     The response to this question is formulated through a detailed analysis of specific applications, which may vary in requirements.
     
    


\begin{table*}[]
\caption{Blockchain-based FL: Test accuracy (\%) on different model combinations - Client A}
\label{tab:proposed-acc-A}
\centering
\def\arraystretch{1.1}
\begin{tabular}{|c|c|cccccccccc|}
\hline
\textbf{Model} &
  \textbf{Params from} &
  \multicolumn{10}{c|}{\textbf{Round}} \\ \hline
\multicolumn{1}{|l|}{} &
  \multicolumn{1}{l|}{} &
  \multicolumn{1}{c|}{\textbf{1}} &
  \multicolumn{1}{c|}{\textbf{2}} &
  \multicolumn{1}{c|}{\textbf{3}} &
  \multicolumn{1}{c|}{\textbf{4}} &
  \multicolumn{1}{c|}{\textbf{5}} &
  \multicolumn{1}{c|}{\textbf{6}} &
  \multicolumn{1}{c|}{\textbf{7}} &
  \multicolumn{1}{c|}{\textbf{8}} &
  \multicolumn{1}{c|}{\textbf{9}} &
  \textbf{10} \\ \hline
\multirow{5}{*}{Simple NN} &
  A &
  \multicolumn{1}{c|}{0.1435} &
  \multicolumn{1}{c|}{0.3143} &
  \multicolumn{1}{c|}{0.3901} &
  \multicolumn{1}{c|}{0.4343} &
  \multicolumn{1}{c|}{\textbf{0.4686}} &
  \multicolumn{1}{c|}{0.4923} &
  \multicolumn{1}{c|}{\textbf{0.5295}} &
  \multicolumn{1}{c|}{0.5405} &
  \multicolumn{1}{c|}{0.559} &
  0.5719 \\ \cline{2-12} 
 &
  A,B &
  \multicolumn{1}{c|}{\textbf{0.1732}} &
  \multicolumn{1}{c|}{\textbf{0.3193}} &
  \multicolumn{1}{c|}{0.3909} &
  \multicolumn{1}{c|}{\textbf{0.4365}} &
  \multicolumn{1}{c|}{0.4666} &
  \multicolumn{1}{c|}{\textbf{0.5027}} &
  \multicolumn{1}{c|}{0.5245} &
  \multicolumn{1}{c|}{0.5465} &
  \multicolumn{1}{c|}{0.567} &
  0.5777 \\ \cline{2-12} 
 &
  A,C &
  \multicolumn{1}{c|}{\textbf{0.1732}} &
  \multicolumn{1}{c|}{\textbf{0.3193}} &
  \multicolumn{1}{c|}{0.3909} &
  \multicolumn{1}{c|}{\textbf{0.4365}} &
  \multicolumn{1}{c|}{0.4666} &
  \multicolumn{1}{c|}{\textbf{0.5027}} &
  \multicolumn{1}{c|}{0.5245} &
  \multicolumn{1}{c|}{0.5465} &
  \multicolumn{1}{c|}{0.567} &
  0.5777 \\ \cline{2-12} 
 &
  B,C &
  \multicolumn{1}{c|}{\textbf{0.1732}} &
  \multicolumn{1}{c|}{\textbf{0.3193}} &
  \multicolumn{1}{c|}{0.3909} &
  \multicolumn{1}{c|}{\textbf{0.4365}} &
  \multicolumn{1}{c|}{0.4666} &
  \multicolumn{1}{c|}{\textbf{0.5027}} &
  \multicolumn{1}{c|}{0.5245} &
  \multicolumn{1}{c|}{0.5465} &
  \multicolumn{1}{c|}{0.567} &
  0.5777 \\ \cline{2-12} 
  &
  A, B, C &
  \multicolumn{1}{c|}{0.1} &
  \multicolumn{1}{c|}{0.3161} &
  \multicolumn{1}{c|}{\textbf{0.3921}} &
  \multicolumn{1}{c|}{0.4348} &
  \multicolumn{1}{c|}{0.4678} &
  \multicolumn{1}{c|}{0.5008} &
  \multicolumn{1}{c|}{0.5287} &
  \multicolumn{1}{c|}{\textbf{0.5503}} &
  \multicolumn{1}{c|}{\textbf{0.5676}} &
   \textbf{0.5815}\\ \hline \hline
\multirow{5}{*}{Efficient-B0} &
  A &
  \multicolumn{1}{c|}{0.7753} &
  \multicolumn{1}{c|}{0.7976} &
  \multicolumn{1}{c|}{0.8121} &
  \multicolumn{1}{c|}{0.8154} &
  \multicolumn{1}{c|}{0.8148} &
  \multicolumn{1}{c|}{0.8279} &
  \multicolumn{1}{c|}{0.8261} &
  \multicolumn{1}{c|}{0.8247} &
  \multicolumn{1}{c|}{0.8315} &
   0.8219\\ \cline{2-12} 
 &
  A,B &
  \multicolumn{1}{c|}{0.7876} &
  \multicolumn{1}{c|}{0.8237} &
  \multicolumn{1}{c|}{0.8391} &
  \multicolumn{1}{c|}{0.843} &
  \multicolumn{1}{c|}{0.843} &
  \multicolumn{1}{c|}{0.848} &
  \multicolumn{1}{c|}{0.8507} &
  \multicolumn{1}{c|}{0.8512} &
  \multicolumn{1}{c|}{0.8483} &
  0.8396 \\ \cline{2-12} 
 &
  A,C &
  \multicolumn{1}{c|}{0.7905} &
  \multicolumn{1}{c|}{0.8222} &
  \multicolumn{1}{c|}{0.8406} &
  \multicolumn{1}{c|}{0.8443} &
  \multicolumn{1}{c|}{0.8441} &
  \multicolumn{1}{c|}{0.8489} &
  \multicolumn{1}{c|}{0.8476} &
  \multicolumn{1}{c|}{0.8495} &
  \multicolumn{1}{c|}{0.8505} &
   0.8449 \\ \cline{2-12} 
 &
  B,C &
  \multicolumn{1}{c|}{\textbf{0.7924}} &
  \multicolumn{1}{c|}{0.8266} &
  \multicolumn{1}{c|}{0.8415} &
  \multicolumn{1}{c|}{0.8399} &
  \multicolumn{1}{c|}{0.8409} &
  \multicolumn{1}{c|}{\textbf{0.8523}} &
  \multicolumn{1}{c|}{0.8493} &
  \multicolumn{1}{c|}{0.8479} &
  \multicolumn{1}{c|}{0.8486} &
  0.8438 \\ \cline{2-12}
  &
  A, B, C &
  \multicolumn{1}{c|}{0.789} &
  \multicolumn{1}{c|}{\textbf{0.8357}} &
  \multicolumn{1}{c|}{\textbf{0.8473}} &
  \multicolumn{1}{c|}{\textbf{0.8492}} &
  \multicolumn{1}{c|}{\textbf{0.8492}} &
  \multicolumn{1}{c|}{0.8521} &
  \multicolumn{1}{c|}{\textbf{0.8534}} &
  \multicolumn{1}{c|}{\textbf{0.8533}} &
  \multicolumn{1}{c|}{\textbf{0.8526}} &
  \textbf{0.8498} \\ \hline 
\end{tabular}
\end{table*}

\begin{table*}[]
\caption{Blockchain-based FL: Test accuracy (\%) on different model combinations - Client B}
\label{tab:proposed-acc-B}
\centering
\def\arraystretch{1.1}
\begin{tabular}{|c|c|cccccccccc|}
\hline
\textbf{Model} &
  \textbf{Params from} &
  \multicolumn{10}{c|}{\textbf{Round}} \\ \hline
\multicolumn{1}{|l|}{} &
  \multicolumn{1}{l|}{} &
  \multicolumn{1}{c|}{\textbf{1}} &
  \multicolumn{1}{c|}{\textbf{2}} &
  \multicolumn{1}{c|}{\textbf{3}} &
  \multicolumn{1}{c|}{\textbf{4}} &
  \multicolumn{1}{c|}{\textbf{5}} &
  \multicolumn{1}{c|}{\textbf{6}} &
  \multicolumn{1}{c|}{\textbf{7}} &
  \multicolumn{1}{c|}{\textbf{8}} &
  \multicolumn{1}{c|}{\textbf{9}} &
  \textbf{10} \\ \hline
\multirow{5}{*}{Simple NN} &
  B &
  \multicolumn{1}{c|}{0.1435} &
  \multicolumn{1}{c|}{0.3143} &
  \multicolumn{1}{c|}{0.3901} &
  \multicolumn{1}{c|}{0.4343} &
  \multicolumn{1}{c|}{\textbf{0.4686}} &
  \multicolumn{1}{c|}{0.4923} &
  \multicolumn{1}{c|}{\textbf{0.5295}} &
  \multicolumn{1}{c|}{0.5405} &
  \multicolumn{1}{c|}{0.559} &
  0.5719 \\ \cline{2-12} 
 &
  B, A &
  \multicolumn{1}{c|}{\textbf{0.1732}} &
  \multicolumn{1}{c|}{\textbf{0.3193}} &
  \multicolumn{1}{c|}{0.3909} &
  \multicolumn{1}{c|}{\textbf{0.4365}} &
  \multicolumn{1}{c|}{0.4666} &
  \multicolumn{1}{c|}{\textbf{0.5027}} &
  \multicolumn{1}{c|}{0.5245} &
  \multicolumn{1}{c|}{0.5465} &
  \multicolumn{1}{c|}{0.567} &
  0.5777 \\ \cline{2-12} 
 &
  B, C &
  \multicolumn{1}{c|}{\textbf{0.1732}} &
  \multicolumn{1}{c|}{\textbf{0.3193}} &
  \multicolumn{1}{c|}{0.3909} &
  \multicolumn{1}{c|}{\textbf{0.4365}} &
  \multicolumn{1}{c|}{0.4666} &
  \multicolumn{1}{c|}{\textbf{0.5027}} &
  \multicolumn{1}{c|}{0.5245} &
  \multicolumn{1}{c|}{0.5465} &
  \multicolumn{1}{c|}{0.567} &
  0.5777 \\ \cline{2-12} 
 &
  A, C &
  \multicolumn{1}{c|}{\textbf{0.1732}} &
  \multicolumn{1}{c|}{\textbf{0.3193}} &
  \multicolumn{1}{c|}{0.3909} &
  \multicolumn{1}{c|}{\textbf{0.4365}} &
  \multicolumn{1}{c|}{0.4666} &
  \multicolumn{1}{c|}{\textbf{0.5027}} &
  \multicolumn{1}{c|}{0.5245} &
  \multicolumn{1}{c|}{0.5465} &
  \multicolumn{1}{c|}{0.567} &
  0.5777 \\ \cline{2-12} 
  &
  A, B, C &
  \multicolumn{1}{c|}{0.1} &
  \multicolumn{1}{c|}{0.3161} &
  \multicolumn{1}{c|}{\textbf{0.3921}} &
  \multicolumn{1}{c|}{0.4348} &
  \multicolumn{1}{c|}{0.4678} &
  \multicolumn{1}{c|}{0.5008} &
  \multicolumn{1}{c|}{0.5287} &
  \multicolumn{1}{c|}{\textbf{0.5503}} &
  \multicolumn{1}{c|}{\textbf{0.5676}} &
   \textbf{0.5815}\\ \hline \hline
\multirow{5}{*}{Eficient-B0} &
  B &
  \multicolumn{1}{c|}{0.7761} &
  \multicolumn{1}{c|}{0.7971} &
  \multicolumn{1}{c|}{0.8121} &
  \multicolumn{1}{c|}{0.8138} &
  \multicolumn{1}{c|}{0.8143} &
  \multicolumn{1}{c|}{0.8258} &
  \multicolumn{1}{c|}{0.8222} &
  \multicolumn{1}{c|}{0.8247} &
  \multicolumn{1}{c|}{0.8292} &
  0.8192 \\ \cline{2-12} 
 &
  B, A &
  \multicolumn{1}{c|}{\textbf{0.7897}} &
  \multicolumn{1}{c|}{0.8281} &
  \multicolumn{1}{c|}{\textbf{0.8408}} &
  \multicolumn{1}{c|}{0.842} &
  \multicolumn{1}{c|}{0.8443} &
  \multicolumn{1}{c|}{0.8503} &
  \multicolumn{1}{c|}{0.849} &
  \multicolumn{1}{c|}{0.8498} &
  \multicolumn{1}{c|}{0.8498} &
  0.8433 \\ \cline{2-12} 
 &
  B, C &
  \multicolumn{1}{c|}{0.7888} &
  \multicolumn{1}{c|}{0.8253} &
  \multicolumn{1}{c|}{0.8394} &
  \multicolumn{1}{c|}{0.8398} &
  \multicolumn{1}{c|}{0.8411} &
  \multicolumn{1}{c|}{0.8501} &
  \multicolumn{1}{c|}{0.8477} &
  \multicolumn{1}{c|}{0.8475} &
  \multicolumn{1}{c|}{0.8494} &
  0.8399 \\ \cline{2-12} 
 &
  A, C &
  \multicolumn{1}{c|}{0.7886} &
  \multicolumn{1}{c|}{0.8252} &
  \multicolumn{1}{c|}{0.8363} &
  \multicolumn{1}{c|}{0.8398} &
  \multicolumn{1}{c|}{0.8432} &
  \multicolumn{1}{c|}{0.8469} &
  \multicolumn{1}{c|}{0.8482} &
  \multicolumn{1}{c|}{0.8466} &
  \multicolumn{1}{c|}{0.8477} &
  0.8439 \\ \cline{2-12}
  &
  A, B, C &
  \multicolumn{1}{c|}{0.7894} &
  \multicolumn{1}{c|}{\textbf{0.8369}} &
  \multicolumn{1}{c|}{0.8369} &
  \multicolumn{1}{c|}{\textbf{0.8495}} &
  \multicolumn{1}{c|}{\textbf{0.8471}} &
  \multicolumn{1}{c|}{\textbf{0.8547}} &
  \multicolumn{1}{c|}{\textbf{0.8549}} &
  \multicolumn{1}{c|}{\textbf{0.8554}} &
  \multicolumn{1}{c|}{\textbf{0.8518}} &
  \textbf{0.8523} \\ \hline 
\end{tabular}
\end{table*}

\begin{table*}[]
\caption{Blockchain-based FL: Test accuracy (\%) on different model combinations - Client C}
\label{tab:proposed-acc-C}
\centering
\def\arraystretch{1.1}
\begin{tabular}{|c|c|cccccccccc|}
\hline
\textbf{Model} &
  \textbf{Params from} &
  \multicolumn{10}{c|}{\textbf{Round}} \\ \hline
\multicolumn{1}{|l|}{} &
  \multicolumn{1}{l|}{} &
  \multicolumn{1}{c|}{\textbf{1}} &
  \multicolumn{1}{c|}{\textbf{2}} &
  \multicolumn{1}{c|}{\textbf{3}} &
  \multicolumn{1}{c|}{\textbf{4}} &
  \multicolumn{1}{c|}{\textbf{5}} &
  \multicolumn{1}{c|}{\textbf{6}} &
  \multicolumn{1}{c|}{\textbf{7}} &
  \multicolumn{1}{c|}{\textbf{8}} &
  \multicolumn{1}{c|}{\textbf{9}} &
  \textbf{10} \\ \hline
\multirow{5}{*}{Simple NN} &
  C &
  \multicolumn{1}{c|}{0.1435} &
  \multicolumn{1}{c|}{0.3143} &
  \multicolumn{1}{c|}{0.3901} &
  \multicolumn{1}{c|}{0.4343} &
  \multicolumn{1}{c|}{\textbf{0.4686}} &
  \multicolumn{1}{c|}{0.4923} &
  \multicolumn{1}{c|}{\textbf{0.5295}} &
  \multicolumn{1}{c|}{0.5405} &
  \multicolumn{1}{c|}{0.559} &
  0.5719 \\ \cline{2-12} 
 &
  C, A &
  \multicolumn{1}{c|}{\textbf{0.1732}} &
  \multicolumn{1}{c|}{\textbf{0.3193}} &
  \multicolumn{1}{c|}{0.3909} &
  \multicolumn{1}{c|}{\textbf{0.4365}} &
  \multicolumn{1}{c|}{0.4666} &
  \multicolumn{1}{c|}{\textbf{0.5027}} &
  \multicolumn{1}{c|}{0.5245} &
  \multicolumn{1}{c|}{0.5465} &
  \multicolumn{1}{c|}{0.567} &
  0.5777 \\ \cline{2-12} 
 &
  C, B &
  \multicolumn{1}{c|}{\textbf{0.1732}} &
  \multicolumn{1}{c|}{\textbf{0.3193}} &
  \multicolumn{1}{c|}{0.3909} &
  \multicolumn{1}{c|}{\textbf{0.4365}} &
  \multicolumn{1}{c|}{0.4666} &
  \multicolumn{1}{c|}{\textbf{0.5027}} &
  \multicolumn{1}{c|}{0.5245} &
  \multicolumn{1}{c|}{0.5465} &
  \multicolumn{1}{c|}{0.567} &
  0.5777 \\ \cline{2-12} 
 &
  A, B &
  \multicolumn{1}{c|}{\textbf{0.1732}} &
  \multicolumn{1}{c|}{\textbf{0.3193}} &
  \multicolumn{1}{c|}{0.3909} &
  \multicolumn{1}{c|}{\textbf{0.4365}} &
  \multicolumn{1}{c|}{0.4666} &
  \multicolumn{1}{c|}{\textbf{0.5027}} &
  \multicolumn{1}{c|}{0.5245} &
  \multicolumn{1}{c|}{0.5465} &
  \multicolumn{1}{c|}{0.567} &
  0.5777 \\ \cline{2-12} 
  &
  A, B, C &
  \multicolumn{1}{c|}{0.1} &
  \multicolumn{1}{c|}{0.3161} &
  \multicolumn{1}{c|}{\textbf{0.3921}} &
  \multicolumn{1}{c|}{0.4348} &
  \multicolumn{1}{c|}{0.4678} &
  \multicolumn{1}{c|}{0.5008} &
  \multicolumn{1}{c|}{0.5287} &
  \multicolumn{1}{c|}{\textbf{0.5503}} &
  \multicolumn{1}{c|}{\textbf{0.5676}} &
   \textbf{0.5815}\\ \hline \hline
\multirow{5}{*}{Efficient-B0} &
  C &
  \multicolumn{1}{c|}{0.7705} &
  \multicolumn{1}{c|}{0.7966} &
  \multicolumn{1}{c|}{0.8142} &
  \multicolumn{1}{c|}{0.8101} &
  \multicolumn{1}{c|}{0.8122} &
  \multicolumn{1}{c|}{0.8306} &
  \multicolumn{1}{c|}{0.8225} &
  \multicolumn{1}{c|}{0.8247} &
  \multicolumn{1}{c|}{0.8275} &
   0.8173 \\ \cline{2-12} 
 &
  C, A &
  \multicolumn{1}{c|}{0.7887} &
  \multicolumn{1}{c|}{0.8236} &
  \multicolumn{1}{c|}{0.8419} &
  \multicolumn{1}{c|}{0.8452} &
  \multicolumn{1}{c|}{0.844} &
  \multicolumn{1}{c|}{0.851} &
  \multicolumn{1}{c|}{0.8448} &
  \multicolumn{1}{c|}{0.8495} &
  \multicolumn{1}{c|}{0.8506} &
   0.8456 \\ \cline{2-12} 
 &
  C, B &
  \multicolumn{1}{c|}{0.7899} &
  \multicolumn{1}{c|}{0.8253} &
  \multicolumn{1}{c|}{0.8377} &
  \multicolumn{1}{c|}{0.8372} &
  \multicolumn{1}{c|}{0.8426} &
  \multicolumn{1}{c|}{0.8502} &
  \multicolumn{1}{c|}{0.8441} &
  \multicolumn{1}{c|}{0.8503} &
  \multicolumn{1}{c|}{\textbf{0.8541}} &
   0.8428 \\ \cline{2-12} 
 &
  A, B &
  \multicolumn{1}{c|}{\textbf{0.7901}} &
  \multicolumn{1}{c|}{0.8233} &
  \multicolumn{1}{c|}{0.8409} &
  \multicolumn{1}{c|}{0.8428} &
  \multicolumn{1}{c|}{0.8456} &
  \multicolumn{1}{c|}{0.848} &
  \multicolumn{1}{c|}{0.8444} &
  \multicolumn{1}{c|}{0.8501} &
  \multicolumn{1}{c|}{0.8509} &
  0.847 \\ \cline{2-12}
  &
  A, B, C &
  \multicolumn{1}{c|}{0.7881} &
  \multicolumn{1}{c|}{\textbf{0.8373}} &
  \multicolumn{1}{c|}{\textbf{0.8429}} &
  \multicolumn{1}{c|}{\textbf{0.8493}} &
  \multicolumn{1}{c|}{\textbf{0.8516}} &
  \multicolumn{1}{c|}{\textbf{0.8564}} &
  \multicolumn{1}{c|}{\textbf{0.8514}} &
  \multicolumn{1}{c|}{\textbf{0.8547}} &
  \multicolumn{1}{c|}{0.8513} &
   \textbf{0.8546}\\ \hline 
\end{tabular}
\end{table*}

\section{Conclusion} \label{sec:conclusion}
This study presents a real-world Ethereum-based FL deployment, encompassing varied testing across various model complexities. Our findings show that the conventional averaging approach on all models, such as Vanilla FL, fails to consistently yield optimal inference results and leads to subpar system performance.  
Instead, empowering clients to select only the appropriate models for their aggregation proved to be a more effective strategy in balancing speed and inference accuracy trade-offs. This, however, is contingent on the complexity of the training model.
For simple and small models, asynchronous aggregation offers a viable and advantageous alternative, eliminating the need to wait for all clients to participate while maintaining the quality of aggregated models.
On the other hand, the selective combinations in the asynchronous aggregation process may not always result in the highest precision in training a large and complex model. Yet, they still can maintain a good aggregated model with only negligible accuracy loss (less than 0.5\%).
Besides, we also observe from our real-world deployment the occurrence of resource exhaustion due to dual tasks on one peer (mining and training model), a scenario that similar research with simulation experiments do not encounter. 

The integration of blockchain technology in our system ensures participants cannot deny their authorship, providing strong evidence against detected abnormal clients. This contributes to the security and integrity of collaborative learning systems, paving the way for more reliable and trustworthy implementations in real-world settings. It is noted that, in the scope of this work, abnormalities do not necessarily imply malicious intent or attempts to disrupt the system; they may arise from the natural data heterogeneity across clients.


Consequently, our future work will focus more on deploying and evaluating the robustness of this method on the non-repudiation in various poisonous data attacks and different datasets. Also, the impact of an arbitrary number of local updates on each peer in asynchronous communication is another intriguing question we aim to explore for optimal values.


\section*{Acknowledgements}
This research is supported by the Research Council of Finland through the 6G Flagship program (Grant 318927), and by Business Finland through the Neural pub/sub research project (8754/31/2022) and Enabling Metaverse (EMETA) (8719/31/2022).


\bibliographystyle{IEEEtran}
\bibliography{ref}
\end{document}